\documentclass[twocolumn,referees]{svjour3}

\usepackage{amsmath}
\usepackage{amssymb}
\usepackage{graphicx}
\usepackage{bm}

\DeclareMathOperator{\sign}{sign}

\begin{document}

\title{Existence of nonlinear normal modes for coupled nonlinear oscillators}

\author{Dirk Hennig}
\institute{Department of Mathematics, University of Portsmouth, Portsmouth, PO1 3HF, UK}

\maketitle

\begin{abstract}
\noindent We prove the existence of nonlinear normal modes 
for general systems of two coupled nonlinear oscillators. 
Facilitating the comparison principle for ordinary differential equations it is shown that there exist exact solutions representing  a 
vibration in unison of the system. The associated  spatially localised time-periodic solutions feature
out-of-phase and in-phase motion of the oscillators.
\end{abstract}

\section{Introduction}

The concept of {\it normal modes} plays a central role for the theory of oscillations of linear system. In a normal mode the dynamics of a 
 finite linear system is equivalent to those of a system with one degree of freedom. 
Linear normal modes (LNMs) can be used to decouple a 
linear system of $n$ degrees of freedom of coupled oscillators such that it performs 
oscillatory motions resulting from the superposition of its eigensolutions with associated eigenfrequencies.
The repercussions are: (i) That if a specific eigenmode only 
is stimulated then the motion is restrained to the 
corresponding harmonic oscillations and energy transfer into other modes is
impossible (invariance). (ii) 
Any form of free or harmonically forced oscillatory
motion can be expressed as the superposition of the 
$n$ modes of natural frequencies. These modes result in stationary periodic 
solutions for which all units
pass through their extreme values of the coordinates and velocities simultaneously (modal superposition). 

However, the assumption of linearity seems too idealised as it is justifiable typically 
for sufficiently small amplitudes only. 
Thus in general one deals inevitably with nonlinear restoring forces for 
which the linear methods such as the superposition principle is not applicable.  

To gain insight into the response of a nonlinear system to excitations 
elaborate methods based on perturbational approaches to derive (approximate) 
solutions have been developed which  rely 
on weak nonlinearity \cite{perturbation}. On the other hand, 
for large amplitude dynamics related to strong nonlinearity the 
expression of exact solutions of the underlying system 
of ordinary differential equations in closed  
analytical form (if existent at all) is in most cases impossible. Regarding exact solutions of finite nonlinear systems the 
concept of nonlinear  normal 
modes (NNMs) was developed to understand dynamical features of systems featuring strong nonlinearity.
Like their linear counterparts NNMs are understood as a {\it vibration in unison} of the system, i.e. 
all units of the system perform synchronous oscillations \cite{Rosenberg}. 
Compared to their linear counterpart for NNMs the superposition principle does not apply 
and the lack of orthogonality  relations restricts their usage as bases for the expression 
of solutions of the underlying nonlinear system in terms of weighted sums of eigenfunctions.
A generalised definition of 
NNMs by geometric means utilising the center manifold technique was given by Shaw and Pierre \cite{Shaw}.  
The existence of NNMs of Hamiltonian systems obeying certain symmetries has been addressed in \cite{existence1}-\cite{existence2}. 

The role of NNMs with regard to the qualitative and quantitative investigation of nonlinear features 
has been illuminated in numerous studies (see e.g. \cite{Rand}-\cite{Belizzi}  and for a recent review  we refer  to
\cite{Avramov} and references therein). To underline the significance of NNMs
for the interpretation of nonlinear phenomena, the  driven resonant motion of nonlinear systems evolves close to NNMs and 
localisation and energy transfer can be explained in terms of NNMs \cite{existence2}. 
Recently the concept of NNMs has been facilitated as a theoretical tool to accomplish targeted 
energy transfer \cite{Kopidakis}-\cite{targeted}. Studies of NNMs in non-smooth systems 
have been performed in \cite{Jiang},\cite{Vestroni}.

The aim of the current paper is to prove the existence of NNMs for a general system of two coupled 
oscillators represented by spatially localised and time-periodic solutions.  
We treat two types of on-site potentials;
namely hard and soft ones.
Unlike for the continuation process of (trivially) localised solutions 
starting from the anti-continuum limit  \cite{Marin}-\cite{Martinez2} 
our approach is not necessarily confined to the weak coupling regime.

\section{The system of two coupled general oscillators}

In this work we prove the existence of spatially localised time-periodic solutions, i.e. NNMs,
for  generic   nonlinear interacting oscillators given by the 
following system
\begin{eqnarray}
\ddot{q}_1(t)&=&-U^{\prime}(q_1(t))-\kappa  V^{\prime}(q_{12}(t)),
\label{eq:start1}\\
\ddot{q}_2(t)&=&-U^{\prime}(q_2(t))+\kappa  V^{\prime}(q_{12}(t)).
\label{eq:start2}
\end{eqnarray}
The variable $q_{n}(t)$ is 
the amplitude of the oscillator at site $n$ evolving in an anharmonic on-site potential 
$U(q_n)$. We introduced the notation $q_{12}=q_1-q_2$.
The prime $^{\prime}$ stands for the derivative with respect to the argument and an overdot $\dot{}$ 
represents the derivative with respect to time $t \in {\mathbb{R}}$. 
The two oscillators interact with each other via an attractive force derived from an  
interaction potential $V(u)$ which 
is analytic and furthermore, is assumed to have the following features:
\begin{equation}
V(0)=V^{\prime}(0)=0,\,\,\, V^{\prime \prime}(0)\ge 0,\,\,\,  V^{\prime \prime}(u \ne 0)> 0.
\label{eq:assumptions1}
\end{equation}
Thus $V(u)$ is convex which is further characterised  by $V^{\prime}(u>0)>0$ and $V^{\prime}(u<0)<0$. 
The interaction potential can be harmonic (diffusive interaction)  but also anharmonic such as e.g. interactions of $\beta-$Fermi-Pasta-Ulam-type 
and Toda-type. The strength of the coupling is determined by the value of $\kappa$.

The on-site potential $U$ is analytic  and is assumed to have the following properties:
\begin{equation}
U(0)=U^{\prime}(0)=0\,,\,\,\,U^{\prime \prime}(0)> 0.\label{eq:assumptions}
\end{equation}
In what follows we differentiate between soft on-site potentials and hard on-site potentials. For the former (latter)
the oscillation frequency of an oscillator moving in the on-site potential $U(q)$
decreases (increases) with increasing oscillation amplitude.
A soft potential possesses at least one inflection point. If a soft potential possesses a 
single inflection point, denoted by $q_i$, 
we suppose without loss of generality  (w.l.o.g.) that $q_i>0$. Then the following relations are valid
\begin{eqnarray}
U^{\prime}(-\infty <q<0)&<&0,\,\,\, U^{\prime}(0<q<q_i)>0,\\
U^{\prime \prime}(q_{i})&=&0,\,\,\,  U^{\prime \prime }(-\infty<q<q_i)>0.
\end{eqnarray} 
If $U(q)$ possesses two inflection points denoted by $q_{i,-}<0$ and $q_{i,+}>0$  it holds that 
\begin{eqnarray}
U^{\prime}(q_{i,-}<q<0)&<&0,\,\,\,  U^{\prime}(0<q<q_{i,+})>0\label{eq:U1prime}\\
U^{\prime \prime}(q_{i,\pm})&=&0,\,\,\,  U^{\prime \prime}(q_{i,-}<q<q_{i,+})>0.\label{eq:U2prime}
\end{eqnarray}
We remark that $U(q)$ can have more than two inflection points (an example is a periodic potential $U(q)=-\cos(q)$). 
However, in the frame of the current study we are only interested in motion 
between the inflection points adjacent to the minimum of $U(q)$ at $q=0$.  
Hence, in the forthcoming we make the {\bf assumption} that for soft on-site potentials the motion at each lattice site $n=1,2$
stays inbetween the inflection points, viz. $q_{i,-}<q_{l}\le q_n(t)\le q_r<q_{i,+}$, where $U(q)$ is convex.

Hard on-site potentials are, in addition to the assumptions in (\ref{eq:assumptions}), 
characterised in their entire range of definition  by 
\begin{equation}
U^{\prime}(q<0)<0,\,\,\,U^{\prime}(q>0)>0,\,\,\,U^{\prime \prime}(q)> 0.\label{eq:assumptionhard} 
\end{equation}
For hard on-site potentials we make the {\bf assumption} that the motion takes place in the range 
$q_{1,2}(t)\in [Q_l,Q_r]$, with $-\infty <Q_l<0$ and $0<Q_r<\infty$ for $t> 0$.

The system (\ref{eq:start1}),(\ref{eq:start2}) possesses an energy integral 
\begin{equation}
 E=\sum_{n=1}^2\left[\frac{1}{2}\dot{q}_n^2+U(q_n) \right]+\kappa V(q_1-q_2).
\end{equation}
There exists a closed maximum equipotential surface $U(q_1)+U(q_2)+\kappa V(q_1-q_2)=E$ bounding all motions and one has
$\dot{q}_1=\dot{q}_2=0$ on this surface. 
\vspace*{1.0cm}

\section{NNMs for coupled nonlinear oscillators in soft on-site potentials}

\noindent In the following we prove the existence of NNMs for 
the system\,(\ref{eq:start1}),(\ref{eq:start2}) with soft on-site potentials.  
In more detail, we show that
there exists a one-parameter
family of spatially localised and time-periodic solutions 
where the amplitude acts as the parameter.
Localisation means that either $|q_1(t)|\ge |q_2(t)|$ or $|q_1(t)|\le |q_2(t)|$ for all $t$, 
and equality holds only at moments 
of time when the amplitudes of two oscillators pass simultaneously through zero. 
The goal is to prove that such localised periodic solutions exist under 
general conditions on $U$ and $V$. 
We emphasise that the existence of localised solutions has to be distinguished 
from the fact that simply due to exchange 
symmetry, $q_1 \leftrightarrow q_2$, 
of the underlying system 
an in-phase mode with equal amplitudes $q_1(t)=q_2(t)$ always exists. Furthermore,  
if, in addition, the potential $U(q)$ possesses the 
spatial reflexion symmetry $U(q)=U(-q)$, 
an out-of-phase mode characterised by  $q_1(t)=-q_2(t)$ is supported.

\vspace*{0.5cm}
\noindent {\bf Theorem 1:} Let $(q_n(t),\dot{q}_n(t))$ be the smooth solutions 
to Eqs.\,(\ref{eq:start1}),(\ref{eq:start2}) with a soft on-site potential satisfying the assumptions above. 
Then there exist periodic solutions $(q_n(t+T_b),\dot{q}_n(t+T_b))=(q_n(t),\dot{q}_n(t))$ for  $n=1,2$,
so that   
the oscillators perform either in-phase motion, i.e. 
$\sign (q_1(t))=\sign (q_{2}(t))$,
 or out-of-phase motion, i.e.
 $\sign (q_1(t))=-\sign (q_{2}(t))$,  
 with period $T_b=2\pi/\omega_b$ and frequency 
$\omega_b$ satisfying
\begin{equation}
\sqrt{\min\{U^{\prime \prime}(q_l),U^{\prime \prime}(q_r)\}}\le \omega_b < 
\sqrt{U^{\prime \prime}(0)+2\tilde{\kappa}}
\label{eq:ineq1}
\end{equation}
where $\tilde{\kappa}=\max\{V^{\prime \prime}(q_l),V^{\prime \prime}(q_r)\} \kappa$.

Moreover, the  solutions are localised which is characterised by either 
\begin{equation}
 |q_1(t)|\ge |q_{2}(t)|,\,\,\,t \in {\mathbb{R}}
\end{equation}
or
\begin{equation}
 |q_2(t)|\ge |q_{1}(t)|,\,\,\,t \in {\mathbb{R}}.
\end{equation}

\vspace*{0.5cm}
\noindent {\bf Proof:} W.l.o.g. the initial conditions satisfy 
\begin{equation}
 q_{1}(0)=q_2(0) =0,\,\,\,  |\dot{q}_{2}(0)| < |\dot{q}_{1}(0)|.\label{eq:ic2}
\end{equation} First, we consider  initial conditions 
$q_{1}(0)=q_2(0)=0$ 
and in-phase initial velocities 
$0<\dot{q}_{2}(0) <\dot{q}_{1}(0)$ and show the existence of localised periodic in-phase solutions. 

Then due to continuity there must exist some 
$t_*>0$ so that during the interval $[0,t_*]$ the 
following  order relation is satisfied
\begin{equation}
 q_{2}(t) \le q_{1}(t).\label{eq:order1}
\end{equation}

We define the difference variable between the coordinates at sites $n=1$ and $n=2$ as follows
\begin{equation}
\Delta q_*(t)=  q_{1}(t)-  q_{2}(t).
\end{equation} 
Thus by definition  $\Delta q_*(t) \ge 0$ on $[0,t_*]$.

The time evolution of the difference variables  $\Delta q_* (t)$ is 
determined by the following equation 
\begin{equation}
 \frac{d^2 \Delta q_*}{dt^2}= -\left[U^{\prime}(q_{1})-U^{\prime}(q_{2})\right]-2\kappa \frac{\partial V(q_1-q_2)}{\partial q_1}
  \label{eq:deltap},
\end{equation}
where we used that ${\partial V(q_{1}-q_{2})}/{\partial q_2}=-{\partial V(q_{1}-q_{2})}/{\partial q_1}$.

Discarding the negative term 
$-2{\kappa}{\partial V(q_{1}-q_{2})}/{\partial q_1}<0$  and utilising that for $q_1(t) \ge q_2(t)$
\begin{equation}
 U^{\prime \prime}(0)(q_1-q_2)\ge U^{\prime}(q_1)-U^{\prime}(q_2)\ge 
 \Omega^2_s (q_1-q_2)>0.
\end{equation}
with $\Omega^2_s=\min\{U^{\prime \prime}(q_l),U^{\prime \prime}(q_r)\}$ \cite{AIP} enables us 
to bound the r.h.s. of Eq.\,(\ref{eq:deltap})  
from above as follows: 
\begin{equation}
\frac{d^2 \Delta q_*}{dt^2}\le -\Omega^2_s\Delta q_*(t).
\end{equation}

Using the properties of the interaction potential (cf. Eqs.\,(\ref{eq:assumptions1})) one gets
$-\max\{V^{\prime \prime}(q_l),V^{\prime \prime}(q_r)\} u \le -V^{\prime}(u)$ for $u\ge 0$, 
so that we bound the r.h.s. of Eq.\,(\ref{eq:deltap})
for $q_1(t) \ge q_2(t)$
from below as follows:
\begin{equation}
 \frac{d^2 \Delta q_*}{dt^2}\ge -(\omega^2_0+2{\tilde{\kappa}})\Delta q_*(t),
\end{equation}
with $\omega^2_0=U^{\prime \prime}(0)$. 

Therefore, by the comparison principle for differential equations, 
$\Delta q_*(t)$ and $\Delta \dot{q}_*(t)$  are bounded from above and below for given initial conditions by the solution of 
\begin{equation}
\frac{d^2 a}{dt^2}=-\Omega^2_s  a
\label{eq:boundabove}
\end{equation}
and 
\begin{equation}
\frac{d^2 b}{dt^2}=-(\omega^2_0 +2{\tilde{\kappa}}) b,
\label{eq:boundbelow}
\end{equation}
respectively, provided $a(t)\ge 0$ and $b(t)\ge 0$.

The solution to Eq.\,(\ref{eq:boundabove}) and (\ref{eq:boundbelow}) with initial conditions 
$(\Delta q_*(0)=0,\Delta \dot{q}_*(0^+)\equiv \Delta \dot{q}_0 \ne  0)$ is given 
by 
\begin{eqnarray}
  a(t)&=& \frac{\Delta \dot{q}_0}{\Omega_s}\sin(\Omega_s t),\\ \label{eq:qabove}
  \dot{a}(t)&=& \Delta \dot{q}_0\cos(\Omega_s t) 
  \label{eq:pabove}
\end{eqnarray}
and 
\begin{eqnarray}
  b(t)&=& \frac{\Delta \dot{q}_0}{\sqrt{\omega^2_0+2{\tilde{\kappa}}}}
  \sin(\sqrt{\omega^2_0+2{\tilde{\kappa}}}\, t),\\ \label{eq:qbelw}
  \dot{b}(t)&=& \Delta \dot{q}_0\cos(\sqrt{\omega^2_0+2{\tilde{\kappa}}}\, t),
  \label{eq:pbelow}
\end{eqnarray}
respectively where
$a(t)\ge 0$ for $0 \le t \le  \pi/\Omega_s$ and $b(t)\ge 0$ 
for $0 \le t \le  \pi/\sqrt{\omega^2_0+2{\tilde{\kappa}}}$.
By $f(\tau_k^{-})$ and  $f(\tau_k^{+})$ the left-sided 
 and right-sided limits of $f(t)$ for $t \rightarrow \tau_k$ are 
 meant, respectively.

Notice that $d^2 \Delta q_*(t)/dt^2\le 0$ on $(0,t_*)$, that is, the acceleration stays non-positive.
Due to the relations $\Delta \dot{q}_0>0$  in conjunction with the lower bound
   $b(t) \le \Delta q_*(t)$  the order relation
as given in (\ref{eq:order1})   is at least maintained on the interval 
$[0,\pi/\sqrt{\omega^2_0+2{\tilde{\kappa}}}]$. Moreover,  
$\Delta q_*(t)$ 
is bound to grow monotonically 
at least during the interval 
$(0,\pi/(2\sqrt{\omega^2_0+2{\tilde{\kappa}}})]$ 
and attains a least maximal value $\Delta \dot{q}_0/\sqrt{\omega^2_0+2{\tilde{\kappa}}}$. 
Furthermore, $\Delta q_*(t)$ cannot return to zero before 
$t=\pi/\sqrt{\omega^2_0+2{\tilde{\kappa}}}$.

From the upper bound $ \Delta q_*(t) \le a(t)$  one infers that $\Delta q_*(t)$
can attain an absolute maximal value 
$\Delta \dot{q}_0/\Omega_s$ but not before $t=\pi/(2\Omega_s)$ and 
$\Delta q_*(t)$ is bound
to return to zero not later than $t=\pi/\Omega_s$. 
Similarly, $\Delta \dot{q}_*(t)$ is bound to decrease monotonically for $0 <t \le  \pi/\Omega_s$ and 
becomes negative
at a time in the interval\\  $(\pi/(2\sqrt{\omega^2_0+2{\tilde{\kappa}}}),\pi/(2\Omega_s))$. 
Moreover, it holds that\\ $\Delta \dot{q}_*(t)\ge -\Delta \dot{q}_0$ for  
$0\le t\le \pi/\sqrt{\omega^2_0+2{\tilde{\kappa}}}$ and $\Delta \dot{q}_*(t)\le -\Delta \dot{q}_0$ 
for  $t\ge \pi/\Omega_s$.

Therefore, by the smooth dependence  of the solutions 
$(\Delta q(t), \Delta \dot{q}(t))$ on the initial values 
 $(q_{1,2}(0),\dot{q}_{1,2}(0))$, to any chosen initial 
condition $q_1(0)=0,\dot{q}_1(0)\ne 0$ and $q_2(0)=0$ there exists a corresponding $\dot{q}_2(0)$ 
(or vice versa) so that one has $\Delta q_*(t_{*})=\Delta q_*(0)=0$ and 
$\Delta \dot{q}_*(t_{*}^-)=-\Delta \dot{q}_0$ with
$t_* \in [\pi/\sqrt{\omega^2_0+2{\tilde{\kappa}}},\pi/\Omega_s]$.
This implies the symmetry  
\begin{eqnarray}
\Delta q_*(t_{*}/2+\tau)&=&\Delta q_*(t_{*}/2-\tau),\label{eq:sym1}\\
-\Delta \dot{q}_*(t_{*}/2+\tau))&=&\Delta \dot{q}_*(t_{*}/2-\tau))\label{eq:sym2}
\end{eqnarray}
with $0<\tau<t_{*}/2$ and $t_{*}/2$ corresponds to the turning point of the motion when
$\Delta q_*$   attain its maximum 
while $\Delta \dot{q}_*$  is zero. In turn, this implies that  the
motion of the two oscillators possesses the symmetry  
\begin{eqnarray}\
q_n(t_{*}/2+\tau)&=&q_n(t_{*}/2-\tau),\,\,\,n=1,2\\ 
\dot{q}_n(t_{*}/2+\tau)&=&\dot{q}_n(t_{*}/2-\tau)),\,\,\,n=1,2,
\end{eqnarray}
with $0\le\tau\le t_{*}/2$ and $t_{*}/2$ corresponds to the turning point of the motion when $q_{1}$ and
$q_2$ 
assume simultaneously their 
respective maxima 
while $\dot{q}_{1}$ and $\dot{q}_2$ pass simultaneously through zero.
Conclusively, on the interval $[0,t_*]$ 
the two  oscillators evolve through half a  cycle of periodic 
in-phase motion, i.e. $\sign (q_1)=\sign (q_{2})$ and
$\sign(\dot{q}_1)=\sign(\dot{q}_{2})$.

At  $t=t_*$, just as at $t=0$,  the  oscillators pass simultaneously through zero coordinate,
corresponding to the minimum position of the on-site potential at $q_n=0$ and the oscillators 
proceed afterwards with negative amplitude, i.e. $q_n(t)<0$ and 
negative velocity, i.e. $\dot{q}_n(t)<0$, $n=1,2$, 
until the next turning point is reached. 

Since $\dot{q}_{1}(t_*)<\dot{q}_{2}(t_*)<0$ then due to continuity there must exist some 
$t_{**}>0$ so that during the interval $[t_*,t_{**}]$  
the following  order relation is satisfied
\begin{equation}
 q_{1}(t) \le q_{2}(t).\label{eq:order3}
\end{equation} 

For $t\ge  t_*$ we consider the
difference variable between the coordinates at sites $n=1$ and $n=2$ 
as follows
\begin{equation}
\Delta q_{**}(t)=  q_{2}(t)-  q_{1}(t).
\end{equation}
Initialising the dynamics accordingly with $\Delta q_{**}(t_*)=\Delta q_{*}=0$, $\Delta \dot{q}_{**}(t_*^+)=\Delta \dot{q}_*(0^+)$ 
and with the arguments given above it follows that $\Delta q_{**}(t)$ and $\Delta \dot{q}_{**}(t)$  exhibit 
 qualitatively the same features as $\Delta q_{*}(t)$ and $\Delta \dot{q}_{*}(t)$ for $0\le t\le t_*$ and lower and upper bounds on 
$\Delta q_{**}(t)$ and $\Delta \dot{q}_{**}(t)$ are derived equivalently to the ones above. In fact, one has 
\begin{eqnarray}
q_n((t_{*}+t_{**})/2+\tau)&=&q_n((t_*+t_{**})/2-\tau),\\
\dot{q}_n((t_*+t_{**})/2+\tau)&=&\dot{q}_n((t_*+t_{**})/2-\tau),
\end{eqnarray}
with $n=1,2$ and  $0 \le \tau\le (t_{**}-t_{*})/2$ and $(t_*+t_{**})/2$ 
corresponds to the turning point of the motion when $q_{1,2}$  attain their 
minima 
while $\dot{q}_{1,2}$ pass through zero.

In particular the time 
 $t_{**}$ at which 
 $\Delta q_{**}(t_{**})=0$  
 and $\Delta \dot{q}_{**}(t_{**}^-)=-\Delta \dot{q}_{**}(t_*^+)$ lies 
 in the range $t_*+\pi/\sqrt{\omega^2_0+2 \tilde{\kappa}} < t_{**}< t_*+\pi/\Omega_s$.   
 The zero of $\Delta q_{**}$ marks the end of a (first) cycle  of duration 
 $2\pi/\sqrt{\omega^2_0+2 \tilde{\kappa}} <T_b=t_{**}<2\pi/\Omega_s$ of 
 maintained localised oscillation throughout 
 of which the order relation $|q_1(t)|\ge |q_{2}(t)|$ is preserved.

Notice that $t_*$ does not necessarily equals  $t_{**}-t_*$ 
when the oscillators perform motion in on-site potentials without 
reflection symmetry, viz. $U(q) \neq U(-q)$.
We remark that the frequency $\omega_b=2\pi/T_b$ 
depends on the amplitude $\bar{q}=\max\{q_l,q_r\}$ and the latter can be chosen 
such that the non-resonance condition 
$m \omega_b(\bar{q}) \ne \omega_0$ 
for all $m \in \mathbb{Z}$ is satisfied.

\vspace{0.5cm}

In relation to the time-periodicity of the dynamics of localised 
 solutions  (NNMs) beyond times $t\ge t_{**}$ 
we  consider intervals
 \begin{equation}
  I_k:=[t_k,t_{k+1}],\,\,\,\mbox{with integer}\,\,\, k\ge 1,\,\,\,t_1=t_{**}\label{eq:interval1}
\end{equation}
 with  
 \begin{equation}
  t_{k+1} =\left\{ \begin{array}{cl} t_k+t_* & \mbox{for}\,\,\,k\,\,\,  \mbox{odd} \\
                               t_k+ t_{**}-t_* & \mbox{for}\,\,\,k\,\,\, \mbox{even}\,. \\
                              \end{array} \right.\label{eq:interval2}
 \end{equation}
Crucially, on each of 
the intervals $I_k$, $ q_n(t)$ and $\dot{q}_n(t)$, $n=1,2$,  
periodically repeat the behaviour of maintained localised oscillations,
described above for the interval $[0,t_*]$ 
for odd $k$ and, $[t_*,t_{**}]$ for even $k$.
Conclusively,  spatially localised and 
time-periodic  solutions (NNMs) result satisfying
\begin{equation}
 |q_1(t)|\ge |q_{2}(t)|,\label{eq:final}
\end{equation}
and $(q_n(t+T_b),\dot{q}_n(t+T_b))=(q_n(t),\dot{q}_n(t))$ for $n=1,2$ with period $T_b=2\pi/\omega_b$ 
where the frequency $\omega_b$ satisfies the relations
\begin{equation}
 \sqrt{\min\{U^{\prime \prime}(q_l),U^{\prime \prime}(q_r)\}}\le \omega_b
 < \sqrt{U^{\prime \prime}(0)+2\tilde{\kappa}},\label{eq:ranges}
\end{equation}
and the out-of-phase NNM is of higher frequency than its in-phase counterpart.

The solutions possess the symmetries
\begin{eqnarray}
 q_n((t_*+(2l+1)t_{**})/2+t)&=&q_n((t_*+(2l+1)t_{**})/2-t),\nonumber\\
 -\dot{q}_n((t_*+(2l+1)t_{**})/2+t)&=&\dot{q}_n((t_*+(2l+1)t_{**})/2-t),\nonumber
\end{eqnarray}
and 
\begin{eqnarray}
 q_n((t_*+lt_{**})/2+t)&=&q_n((t_*+lt_{**})/2-t),\nonumber\\
 -\dot{q}_n((t_*+lt_{**})/2+t)&=&\dot{q}_n((t_*+lt_{**})/2-t),\nonumber.
\end{eqnarray}
with $n=1,2$, $l \in {\mathbb{Z}}$ and initialising the dynamics with 
\begin{equation}
q_{1,2}(0)=q_{1,2}((t_*+(2l+1)t_{**})/2),\dot{q}_{1,2}(0)=0,\nonumber
\end{equation}
or
\begin{equation}
q_{1,2}(0)=q_{1,2}((t_*+lt_{**})/2),\dot{q}_{1,2}(0)=0\nonumber
\end{equation}
yields time-reversible solutions, viz. $q_{1,2}(t)=q_{1,2}(-t)$ and $-\dot{q}_{1,2}(t)=\dot{q}_{1,2}(-t)$.

Consequently, the relation (\ref{eq:final}) is true for $t \in {\mathbb{R}}$. 
We remark that the case   of out-of-phase initial velocities, 
$\sign(\dot{q}_1(0))=-\sign(\dot{q}_{2}(0))$, 
is treated in the same way as above and the proof is complete.

\hspace{7.5cm} $\square$

\begin{figure}
\includegraphics[scale=0.6]{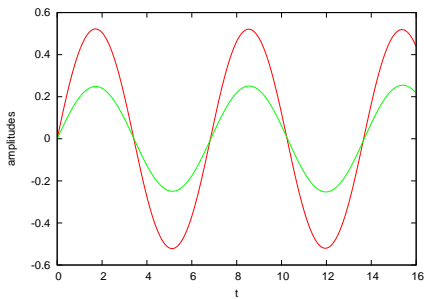}
\caption{Periodic oscillations of the two coordinates $q_1$ 
and $q_2$ with $|q_1(t)|\ge |q_2(t)|$, corresponding to an in-phase NNM for motion in the soft on-site 
potential given in (\ref{eq:soft}) and harmonic coupling of strength $\kappa=0.1$.
The initial conditions are given by $q_1(0)=q_2(0)=0$ and $p_1(0)=0.5$, $p_2(0)=0.23$.} 
\label{fig:amplitudes}
\end{figure}

For an illustration we show in Fig.~\ref{fig:amplitudes} the periodic in-phase oscillations
of the coordinates $q_1$ and $q_2$ for motion in a soft on-site potential given by
\begin{equation}
 U(q)=\frac{1}{2}q^2-\frac{1}{4}q^4,\label{eq:soft}
\end{equation}
and linear coupling originating from the harmonic interaction potential
\begin{equation}
 V(q_1-q_2)=\frac{\kappa}{2}(q_1-q_2)^2.
\end{equation}
The localisation feature of the NNM is reflected in $|q_1(t)|\ge |q_2(t)|$.

\vspace*{1.0cm}

\section{NNMs for motions in hard on-site potentials}

In this section we consider hard on-site potentials.
The next Theorem establishes the existence of NNMs represented by spatially localised and 
time-periodic solutions in hard on-site potentials.

\vspace*{0.5cm}
\noindent {\bf Theorem 2:} 
Let $(q_n(t),\dot{q}_n(t))$ be the smooth solutions 
to Eqs.\,(\ref{eq:start1}),(\ref{eq:start2}) with a hard on-site
potential satisfying the assumptions above. 
Then there exist periodic solutions $(q_n(t+T_b),\dot{q}_n(t+T_b))=(q_n(t),\dot{q}_n(t))$ for  $n=1,2$,
so that   
the oscillators perform either in-phase motion, i.e. 
$\sign (q_1(t))=\sign (q_{2}(t))$,
 or out-of-phase motion, i.e.
 $\sign (q_1(t))=-\sign (q_{2}(t))$,
with period 
$T_b=2\pi/\omega_b$ and frequency $\omega_b$ satisfying
\begin{equation}
 \sqrt{U^{\prime \prime}(0)}
 < \omega_b\le \sqrt{\max\{U^{\prime \prime}(Q_l),U^{\prime \prime}(Q_r)\}+2\tilde{\kappa}} 
 \label{eq:ineq2}
\end{equation}
where $\tilde{\kappa}=\max\{V^{\prime \prime}(Q_l),V^{\prime \prime}(Q_r)\} \kappa$.
Moreover, the solutions are  localised  fulfilling either 
\begin{equation}
 |q_1(t)|\ge |q_{2}(t)|,\,\,\,t\in {\mathbb{R}}
\end{equation}
or 
\begin{equation}
 |q_2(t)|\ge |q_{1}(t)|,\,\,\,t\in {\mathbb{R}}
\end{equation}

\vspace*{0.5cm}
\noindent {\bf Proof:} W.l.o.g. the initial conditions satisfy 
\begin{equation}
 q_{1}(0)=q_2(0) =0,\,\,\,  |\dot{q}_{2}(0)| < |\dot{q}_{1}(0)|.
\end{equation} First, we consider  initial conditions 
$q_{1}(0)=q_2(0)=0$ 
and in-phase initial velocities 
$0<\dot{q}_{2}(0) <\dot{q}_{1}(0)$ and show the existence of localised periodic in-phase solutions. 
(The treatment of out-of-phase initial velocities, 
$\sign(\dot{q}_1(0))=-\sign(\dot{q}_{2}(0))$,  
proceeds in the same manner.)
Due to continuity there must exist some 
$t_*>0$ so that during the interval $[0,t_*]$ the two oscillators  perform 
motion with $\sign (q_1(t))=\sign (q_{2}(t))$ and
$\sign(\dot{q}_1(t))=\sign(\dot{q}_{2}(t))$.
Furthermore, the 
following  order relation is satisfied
\begin{equation}
 q_{2}(t) \le q_{1}(t).\label{eq:order2}
\end{equation} 

We proceed as in the previous case for soft on-site potentials by introducing the difference variable between 
coordinates. The time evolution of the difference variable is determined by an equation identical to 
(\ref{eq:deltap}) and using 
\begin{equation}
 \Omega_h^2(q_1-q_2)\ge U^{\prime}(q_1)-U^{\prime}(q_2)\ge 
 U^{\prime \prime}(0) (q_1-q_2)>0
\end{equation}
\cite{AIP} and
$-\max\{V^{\prime \prime}(Q_l),V^{\prime \prime}(Q_r)\} u \le -V^{\prime}(u)$ for $u\ge 0$, 
we derive for $q_1(t)\ge q_2(t)$  for the r.h.s. an 
upper bound and lower bound for the r.h.s. for hard on-site potentials as
 \begin{equation}
\frac{d^2 \Delta q_*}{dt^2}\le -\omega^2_0\Delta q_*(t),
\end{equation}
 and 
\begin{equation}
 \frac{d^2 \Delta q_*}{dt^2}\ge  -(\Omega^2_h+2\tilde{\kappa})\Delta q_*(t),
\end{equation}
respectively and $\Omega^2_h=\max\{U^{\prime \prime}(Q_l),U^{\prime \prime}(Q_r)\}$.

Thus, the solutions are bounded from above and below as $B(t)\le \Delta q_*(t)\le A(t)$ 
where the upper bound is given by 
\begin{equation}
   A(t)= \frac{\Delta \dot{q}_0(0)}{\sqrt{\Omega_h^2+2\tilde{\kappa}}}
   \sin(\sqrt{\Omega_h^2+2\tilde{\kappa}} t), \label{eq:qaboveh}
 \end{equation}
 and $0 < t <\pi/\sqrt{\Omega_h^2+2\tilde{\kappa}}$.
 \begin{equation}
   B(t)= \frac{\Delta \dot{q}_0(0)}{\omega_0}
   \sin(\omega_0\, t), \label{eq:qbelowh}
 \end{equation}
and $0 <t <\pi/\omega_0$.

The remainder of the proof regarding the time periodicity of 
the localised solutions proceeds 
in an analogous way as above for Theorem 1.

Conclusively,  spatially localised and time-periodic solutions (NNMs)
for out-of-phase motion in hard on-site potentials result which satisfy
\begin{equation}
 |q_1(t)|\ge |q_{2}(t)|,\,\,\,t \ge 0,
\end{equation}
and $(q_n(t+T_b),\dot{q}_n(t+T_b))=(q_n(t),\dot{q}_n(t))$ for $n=1,2$ and with period
$T_b=2\pi /\omega_b$. 
The frequencies lie in the interval
\begin{equation}
 \sqrt{U^{\prime \prime}(0)} < \omega_b \le 
 \sqrt{\max\{U^{\prime \prime}(Q_l),U^{\prime \prime}(Q_r)\}+2\tilde{\kappa}}
\end{equation}
and the out-of-phase NNM is of higher frequency than its in-phase counterpart.
The amplitude $\bar{Q}=\max\{Q_l,Q_r\}$, can be chosen such that the non-resonance 
condition $m \omega_b(\bar{Q}) \ne \omega_0$ for all $m \in \mathbb{Z}$ is satisfied 
completing the proof.

\hspace{7.5cm} $\square$

\section{Summary} 
 
We have proven the existence of exact  time-periodic spatially localised solutions, 
i.e. localised NNMs, for two coupled general nonlinear oscillators
utilising the comparison principle for ODEs. In more detail,  
in systems with an anharmonic on-site 
potential $U$ the existence of in-phase and out-of-phase 
localised periodic solutions has been 
proven. Furthermore, suppose the interaction potential $V(q)$ possesses 
the property $V^{\prime \prime}(0)>0$ so that for small arguments the harmonic limit of $V(q)$ is valid. 
Then, when the amplitude of the NNMs tends to zero the linear NMs of two linearly 
coupled harmonic oscillators are recovered, viz. a mode 
of in-phase oscillations ($q_1(t)=q_2(t)$) and and a mode of out-of-phase 
 oscillations ($q_1(t)=-q_2(t)$) of the two oscillators with  
  frequency $\omega_0$  and $\sqrt{\omega_0^2+2\kappa}$ respectively. 

The localised NNMs as discussed above and, in general, 
  equal-amplitude NNMs and LNMs, 
  have in common that they are characterised by a 
  vibration in unison of the system (their involved units pass
  through their extreme values of the coordinates and velocities 
  simultaneously). In this context, the localised solutions 
  to the system of two coupled oscillators resemble 
also the localisation behaviour exhibited by breather solutions in extended lattice 
systems where only a few oscillators oscillate with
considerable amplitude while the others 
oscillate with much smaller amplitudes. 

Our developed method is expected to stimulate further research regarding 
the existence of time-periodic space-localised patterns and their formation in  extended 
networks of generic coupled nonlinear oscillators.  In particular, the method developed in this paper
can be utilised  to prove the existence of NNMs in finite size lattices with global coupling.

\end{document}